\documentclass[9pt,conference]{IEEEtran}
\IEEEoverridecommandlockouts

\usepackage{cite}
\usepackage{amsmath,amssymb,amsfonts}
\usepackage{algorithmic}
\usepackage{graphicx}
\usepackage{textcomp}
\usepackage{xcolor}
\usepackage{hyperref}

\usepackage{multirow}  
\usepackage{array}    
\usepackage{booktabs}  
\usepackage{makecell}  
\def\BibTeX{{\rm B\kern-.05em{\sc i\kern-.025em b}\kern-.08em
    T\kern-.1667em\lower.7ex\hbox{E}\kern-.125emX}}
\begin{document}

\title{Noise Supervised Contrastive Learning and Feature-Perturbed for Anomalous Sound Detection\\

}


\author{
  \IEEEauthorblockN{Shun Huang}
  \IEEEauthorblockA{
    \textit{School of Computer Science and Technology} \\
    \textit{Xinjiang University} \\
    \textit{Urumqi, China} \\
    huangswt@stu.xju.edu.cn
  }
  \and
  \IEEEauthorblockN{Zhihua Fang}
  \IEEEauthorblockA{
    \textit{School of Computer Science and Technology} \\
    \textit{Xinjiang University} \\
    \textit{Urumqi, China} \\
    fangzhihua@stu.xju.edu.cn
  }
  \and
  \IEEEauthorblockN{Liang He\IEEEauthorrefmark{1}\thanks{\IEEEauthorrefmark{1}Corresponding Author}\thanks{This work was supported in part by the National Natural Science Foundation of China under Grant 62366051, and in part by the State Grid Xinjiang Electric Power Company and Xinjiang Siji Information Technology Co., Ltd. under Grant SGITXX00ZHXX2200262.}}
  \IEEEauthorblockA{
    \textit{Department of Electronic Engineering} \\
    \textit{Tsinghua University} \\
    \textit{Beijing, China} \\
    heliang@mail.tsinghua.edu.cn
  }
}


\maketitle

\begin{abstract}

Unsupervised anomalous sound detection aims to detect unknown anomalous sounds by training a model using only normal audio data.  Despite advancements in self-supervised methods, the issue of frequent false alarms when handling samples of the same type from different machines remains unresolved.  This paper introduces a novel training technique called one-stage supervised contrastive learning (OS-SCL), which significantly addresses this problem by perturbing features in the embedding space and employing a one-stage noisy supervised contrastive learning approach.  On the DCASE 2020 Challenge Task 2, it achieved 94.64\% AUC, 88.42\% pAUC, and 89.24\% mAUC using only Log-Mel features.  Additionally, a time-frequency feature named TFgram is proposed, which is extracted from raw audio.  This feature effectively captures critical information for anomalous sound detection, ultimately achieving 95.71\% AUC, 90.23\% pAUC, and 91.23\% mAUC.  The source code is available at: \underline{www.github.com/huangswt/OS-SCL}. 

\end{abstract}


\section{Introduction}
Anomalous sound detection (ASD) technology is increasingly applied in the industrial sector, driving the development of new techniques. However, due to the scarcity of anomalous audio samples in real environments, the specifics of machine malfunctions are often unknown, and machines typically operate normally. Thus, anomalous sound detection is defined as an unsupervised problem.

Existing anomalous sound detection systems are primarily divided into two categories: methods based on autoencoder reconstruction \cite{suefusa2020anomalous,Daniluk2020,Hayashi2020} and those using machine ID self-supervised \cite{giri2020self,liu2022anomalous,zhang23fa_interspeech,guan2023anomalous,10447764} classification models . Autoencoder-based (AE) systems use reconstruction error as an anomaly score to detect anomalies. For some non-stationary machine sounds, even without considering anomalies, the reconstruction error for edge frames is often large. Interpolation deep neural networks (IDNN) \cite{suefusa2020anomalous} address this by removing the middle frame as input and predicting the interpolated frame as output, effectively handling non-stationary machine detection. Due to the good generalization of AE \cite{10096813}, anomalous segments might still be reconstructed successfully, leading to smaller anomaly scores, even without anomalous data during training. In Glow-Aff \cite{dohi2021flow}, normalized flow is used to estimate the probability density of normal data to mitigate overfitting, but it requires training a separate model for each machine ID. Due to the unknown nature of anomalous audio, modeling a stable distribution is often challenging \cite{8501554}.

\begin{figure*}[htbp]
\centerline{\includegraphics[width=1\textwidth]{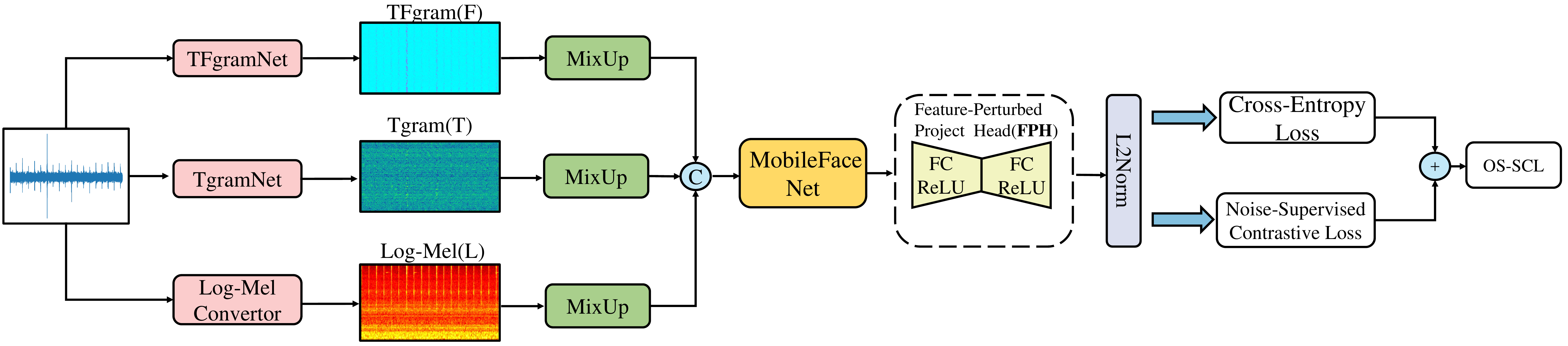}}
\caption{Framework of One-Stage Noise-Supervised Contrastive Learning with Embedding Space Feature Perturbation.}
\label{fig}
\end{figure*}

The machine ID-based self-supervised classification model is a promising approach, consistently outperforming autoencoder-based reconstruction methods in detecting anomalies across multiple machines. Previous MobileNet V2 classification models \cite{giri2020self} have shown good performance in anomalous sound detection, but there is significant variability in detection performance across different IDs of the same machine type, indicating instability. To address the potential lack of high-frequency components in the Log-Mel spectrogram, STgram \cite{liu2022anomalous} uses a convolutional neural network (TgramNet) to directly extract temporal information from raw audio to supplement the Log-Mel spectrogram. ASD-AFPA \cite{zhang23fa_interspeech} employs a multi-head self-attention mechanism (MHSA) \cite{DBLP:conf/nips/VaswaniSPUJGKP17} to address the automatic identification of frequency patterns in machine sounds. CLP-SCF \cite{guan2023anomalous} employs a two-stage contrastive learning \cite{NEURIPS2020_d89a66c7} and self-supervised classification fine-tuning to reduce feature embedding distances from the same machine ID and increase distances from other machine IDs. Noisy-ArcMix \cite{10447764} uses MixUp \cite{zhang2018mixup} to construct a noise-infused loss function to enhance generalization capability. However, these methods still do not solve the core issue in ID self-supervised classification: when normal samples from different machine IDs are very similar, it is difficult to train clear decision boundaries, leading to frequent false alarms \cite{Koizumi_DCASE2020_01}. This is also why these anomalous sound detection systems perform poorly on the Toy-conveyor dataset. Although large pre-trained models \cite{jiang24c_interspeech,10447183,LvHUAKONG2023} have demonstrated excellent performance in anomalous sound detection, their large parameter size poses challenges for deployment in real industrial environments, and inference delays remain a significant bottleneck. Finally, it is commonly believed that the Log-Mel spectrogram may filter out high-frequency components of anomalous audio \cite{liu2022anomalous}, and that machine anomaly detection relies on high-frequency components.  However, this is a misconception.

To address these issues, we propose a novel training technique called One-Stage Supervised Contrastive Learning (OS-SCL). First, we apply MixUp \cite{zhang2018mixup} to blend data within the same batch, where different mixing ratios result in features similar to either original or shuffled features. In the embedding space, we construct a Feature Perturbation Head (FPH) to apply perturbations to the embedded features and simultaneously calculate both the supervised contrastive loss  and cross-entropy loss to optimize the model. Additionally, inspired by the PANN \cite{9229505} framework, we modify the original network structure by incorporating a global max pooling layer, enabling the network to more comprehensively capture features in audio signals across both the time and frequency domains. We refer to this feature as TFgram. Finally, our detection performance using only the Log-Mel spectrogram significantly outperforms STgram \cite{liu2022anomalous}, proving that machine anomaly sound detection does not rely on high-frequency components. Our contributions can be summarized as follows:
\begin{itemize}
    \item We propose the one-stage noise-supervised contrastive learning (OS-SCL) training technique, which significantly reduces the false alarm rate in self-supervised classification compared to the latest methods and achieves the best performance in the DCASE2020 Challenge.
    \item We demonstrate that machine anomaly sound detection is not sensitive to high-frequency components, while achieving higher detection stability.
    \item Our method requires significantly fewer parameters than large pre-trained models, yet it still provides comparable anomaly detection performance.

\end{itemize}

\section{Proposed Method}
As shown in Fig.~\ref{fig}, the overall framework of OS-SCL includes three key components: feature perturbation mapping head for perturbing features in the embedding space, noise-supervised contrastive learning for optimizing embedded features, and cross-entropy feature classification.  Additionally, it includes a time-frequency feature extracted from the raw audio for anomalous sound detection.  This section will provide a detailed introduction to each of these components.

\subsection{Feature Mapping Perturbation}
Let a batch of single-channel audio signals be \( x \in \mathbb{R}^{B \times L} \), with the corresponding log-Mel spectrogram \( x_{mel} \in \mathbb{R}^{B \times M \times N} \), where \( B \) denotes the batch size, \( L \) is the length of each audio signal, \( M \) represents the number of Mel frequency filters, and \( N \) is the number of time frames. Mixup \cite{zhang2018mixup} is applied to the audio signal \( x \) and its corresponding log-Mel spectrogram using a mixing coefficient \( \lambda \). The original labels are \( y_a \), and the shuffled labels are \( y_b \), obtained by randomly shuffling \( y_a \). The mixed audio signal and log-Mel spectrogram are denoted as \( x^{i,j} \) and \( x_{mel}^{i,j} \), respectively. After feature embedding through the MobileFaceNet \cite{chen2018mobilefacenets} model, the signal features are mapped to a new representation space, denoted as \( x_{emb}^{i,j} \in \mathbb{R}^{B \times E} \), where \( E \) is the dimension of the embedded features.

We designed a Feature Perturbation Head (FPH), inspired by the autoencoder structure \cite{Harada_arXiv2023_01}, using only the first layer of the encoder to reduce to a compressed dimension \( R \), and restoring through the last layer of the decoder. Although similar in structure to an autoencoder, we do not use reconstruction loss, but focus on feature perturbation. The perturbed features are located on a unit sphere after $L_2$ normalization, eliminating differences in feature value ranges and allowing the model to focus more on feature direction rather than absolute magnitude.

\subsection{Decision Boundary Learning}
Traditional two-stage training methods \cite{guan2023anomalous} often diminish the effect of supervised contrastive learning during fine-tuning stage and overfit the training set, failing to adequately address the fundamental issues of self-supervised classification. Therefore, we adopt a one-stage training technique that completes feature optimization and classification with supervised contrastive learning in a single training process. Additionally, during training, the calculation of supervised contrastive loss consistently uses the original label \( y_a \). However, when the mixup ratio \( \lambda \) is small, the correct label for the mixed embedding feature \( x_{emb}^{i,j} \) is closer to the shuffled label \( y_b \). This label mismatch introduces a certain amount of noise, which actually helps the model learn a more stable decision boundary, effectively distinguishing similar samples of different machine IDs.

Specifically, for \( x_{emb}^{i,j} \), which represents multiple samples embedded from different machine IDs after feature perturbation, we define \( I \) as the set of all samples of \( x_{emb}^{i,j} \), i.e., \( i \in I \equiv \{1...B\} \), where \( B \) is the batch size. For a selected anchor point \( \boldsymbol{z}_i \), \( P(i) \) denotes the set of remaining samples that share the same machine ID with the anchor, and \( A(i) \equiv I \setminus \{i\} \) represents all remaining samples excluding the anchor \( \boldsymbol{z}_i \). The relationship between \( P(i) \) and \( A(i) \) is expressed as \( P(i) \subseteq A(i) \). The embedding features are optimized through the following noise-supervised contrastive learning loss function:

\begin{equation}
\mathcal{L}_{sup}^{Noise}  = \sum_{i \in I} \frac{-1}{|P(i)|} \sum_{p \in P(i)} \log \frac{\exp\left(\boldsymbol{z}_i \cdot \boldsymbol{z}_p/\tau\right)}{\sum_{a \in A(i)} \exp\left(\boldsymbol{z}_i \cdot \boldsymbol{z}_a/\tau\right)}
\end{equation}
where \( i \) represents the current anchor, \( |P(i)| \) indicates the cardinality of the sample set \( P(i) \), and \( \tau \) is the temperature coefficient. In the above supervised contrastive learning loss function, the anchor \( \boldsymbol{z}_i \) forms a positive contrast with positive samples in the set \( P(i) \) (i.e., samples with the same machine ID as \( \boldsymbol{z}_i \)), aiming to enhance the cohesion of similar samples by maximizing the similarity between positive samples. Meanwhile, the anchor \( \boldsymbol{z}_i \) contrasts with all samples in the set \( A(i) \), aiming to minimize the similarity between negative samples (i.e., samples with different machine IDs), thereby pushing the distance between different classes of samples.

It is important to note that the calculation of the noise-supervised contrastive loss function consistently uses the original label \( y_a \), meaning that the samples in \( P(i) \) are selected based on \( y_a \). When the mixup coefficient \( \lambda \) is small, the label of the mixed sample is closer to the shuffled label \( y_b \), but the label used for loss calculation remains \( y_a \). This discrepancy may cause some samples in \( P(i) \) to have machine IDs inconsistent with the anchor \( \boldsymbol{z}_i \), thereby introducing label noise. Label noise encourages the model to not only rely on labels but also to explore the deeper features of samples, enhancing sensitivity to subtle differences and improving feature discrimination. Furthermore, label noise helps reduce overfitting and optimizes the decision boundary.

\subsection{Feature Classification}
ArcFace \cite{deng2019arcface} effectively enhances intra-class compactness and inter-class separability by introducing an additive angular margin. However, when the model encounters unseen anomalous samples, ArcFace may not ensure a sufficiently large angular difference between normal and anomalous samples. This limitation can easily lead to confusion between normal and anomalous samples during the testing phase. To address this issue, we introduce the Noisy-ArcMix \cite{10447764} mode, expressed as follows:
\begin{equation}
\begin{aligned}
    \mathcal{L}_{\mathrm{NAMix}} &= \lambda \cdot \mathcal{L}_{\mathrm{CE}}(\mathrm{ArcFace}(x_{emb}^{i,j}, \mathbf{y}_{a}), \mathbf{y}_a) \\
    &\quad + (1 - \lambda) \cdot \mathcal{L}_{\mathrm{CE}}(\mathrm{ArcFace}(x_{emb}^{i,j}, \mathbf{y}_{a}), \mathbf{y}_b)
\end{aligned}
\end{equation}
where \(\lambda\) is the Mixup coefficient, \(\mathcal{L}_{\mathrm{CE}}\) denotes the cross-entropy loss function, \(\mathbf{y}_a\) is the original label, and \(\mathbf{y}_b\) is the label generated by mixing \(\mathbf{y}_a\). The Noisy-ArcMix loss function dynamically adjusts the sensitivity of classification by altering the weight of the mixed labels, thereby enhancing sensitivity to small perturbations.

\subsection{TFgramNet Architecture}
\begin{table}[tp]
  \centering
  \caption{For each layer within the TFgramNET architecture, the symbols C, K, S, P, D, and OS represent the channels, kernel size, stride, padding, dilation rate, and output size respectively.}
  \label{CAFPAFNET_Table}
  \begin{tabular}{ccccccc}
    \toprule
    Layer & C & K & S & P & D & OS \\
    \midrule
    Conv1D & 64 & 11 & 5 & 5 & - & - \\
    BatchNorm+ReLU & - & - & - & - & - & - \\
    Convblock & 64 & 3 & 1 & 1\textbar2 & -\textbar2&- \\
    MaxPool1D & - & 4 & - & - & - & - \\
    \midrule
    Convblock & 64 & 3 & 1 & 1\textbar2 & -\textbar2&- \\
    Adaptive\_MaxPool1D & - & - & - & - & - & 626 \\
    Convblock & 128 & 3 & 1 & 1\textbar2 &  -\textbar2&- \\
    Adaptive\_MaxPool1D & - & - & - & - & - & 313 \\
    \bottomrule
  \end{tabular}
\end{table}

TFgramNet is a modification of PANN \cite{9229505} with global max pooling, designed to learn time-frequency representations, as detailed in Table~\ref{CAFPAFNET_Table}. The extracted TFgram features possess both time and frequency axes. A 1D convolutional layer with a kernel size of 11 and a stride of 5 is first applied to the time-domain mixed waveform \({x}^{i,j}\), outputting a size represented as $F \times T$, where $F$ is the learnable frequency information and $T$ represents the time frames. This is followed by three convolutional blocks, each containing two convolutional layers, with batch normalization \cite{10.5555/3045118.3045167} and ReLU \cite{relu} following each convolutional layer. The first block followed by a max pooling layer and the latter two blocks each followed by a global max pooling layer. Downsampling is performed on the time steps, and the selection of frequency information is continuously optimized during training. The final TFgram obtained can be combined with STgram \cite{liu2022anomalous} for anomalous sound detection, represented as:
\begin{equation}
\mathbf{x}_{TFST}=Concat(\mathbf{x}_{mel}^{i,j},\mathbf{x}_{T}^{i,j},\mathbf{x}_{TF}^{i,j})
\end{equation}
where \(\mathbf{x}_{T}^{i,j}\) denotes the time information Tgram, and \(\mathbf{x}_{TF}^{i,j}\) represents the TFgram with time and frequency axes. Mobile-FaceNet \cite{chen2018mobilefacenets} is chosen as the embedding model for the input features.

Finally, our OS-SCL framework loss function can be expressed as:
\begin{equation}
\mathcal{L}_{OS-SCL} =\mathcal{L}_{sup}^{Noise} +\mathcal{L}_{\mathrm{NAMix}}
\end{equation}

Noise supervised contrastive learning effectively distinguishes similar and different samples in the feature space, enhancing the model classification capability for known categories and providing better generalization performance when handling unseen anomalous samples. The Noisy-ArcMix further enhances anomaly detection accuracy by dynamically adjusting the weight of mixed labels, enabling the model to more acutely recognize subtle perturbations.

\section{Experimental Results}

\begin{table*}[!ht]
    \setlength{\abovecaptionskip}{0cm}
    \centering
    \caption{Performance comparison in terms of AUC (\%) and pAUC (\%) on the test data of the development dataset.}
    \resizebox{0.93\textwidth}{!}{
    \begin{tabular}{ccccccccccccccc}
        \toprule
        \multirow{2}{*}{Methods} & \multicolumn{2}{c}{Fan} & \multicolumn{2}{c}{Pump} & \multicolumn{2}{c}{Slider} & \multicolumn{2}{c}{Valve} & \multicolumn{2}{c}{ToyCar} & \multicolumn{2}{c}{ToyConveyor} & \multicolumn{2}{c}{Average} \\
        \cmidrule(lr){2-3} \cmidrule(lr){4-5} \cmidrule(lr){6-7} \cmidrule(lr){8-9} \cmidrule(lr){10-11} \cmidrule(lr){12-13} \cmidrule(lr){14-15} 
         & AUC & pAUC & AUC & pAUC & AUC & pAUC & AUC & pAUC & AUC & pAUC & AUC & pAUC & AUC & pAUC \\
        \midrule
        IDNN \cite{suefusa2020anomalous}  & 67.71 & 52.90 & 73.76 & 61.07 & 86.45 & 67.58 & 84.09 & 64.94 & 78.69 & 69.22 & 71.07 & 59.70 & 76.96 & 62.57 \\
        MobileNetV2 \cite{giri2020self}  & 80.19 & 74.40 & 82.53 & 76.50 & 95.27 & 85.22 & 88.65 & 87.98 & 87.66 & 85.92 & 69.71 & 56.43 & 84.34 & 77.74 \\
        Glow\_Aff \cite{dohi2021flow}  & 74.90 & 65.30 & 83.40 & 73.80 & 94.60 & 82.80 & 91.40 & 75.00 & 92.20 & 84.10 & 71.50 & 59.00 & 85.20 & 73.90 \\
        
        STgram-MFN \cite{liu2022anomalous}  & 89.31 & 83.71 & 93.54 & 85.08 & 99.58 & 97.82 & 99.18 & 96.45 & 95.15 & 90.06 & 73.62 & 62.01 & 91.73 & 85.86 \\
    
        CLP-SCF \cite{guan2023anomalous} & 96.98 & 93.23 & 94.97 & 87.39 & 99.57 & 97.73 & 99.89 & 99.51 & 95.85 & 90.19 & 75.21 & 62.79 & 93.75 & 88.48 \\

        TASTgram(NAMix) \cite{10447764} & 98.10 & \textbf{94.97} & 94.51 & 85.22 & 99.49 & 97.30 & \textbf{99.95} & \textbf{99.72} & 96.19 & 89.96 & 76.26 & 66.95 & 94.08 & 89.01
 \\
        \midrule
        \textbf{Log-Mel(OS-SCL)} & 95.66 & 88.59 & 94.34 & 85.80 & 99.55 & 97.64 & 99.83 & 99.09 & \textbf{96.83} & 90.44 & 81.64 & 68.97 & 94.64 & 88.42 \\
        \textbf{TFSTgram(OS-SCL)} & \textbf{98.41} & 94.50 & \textbf{96.25} & \textbf{88.74} & \textbf{99.63} & \textbf{98.04} & 99.85 & 99.23 & 96.55 & \textbf{91.17} & \textbf{83.56} & \textbf{69.87} & \textbf{95.71} & \textbf{90.23}
        
        \\

        \bottomrule
    \end{tabular}
    }
    \label{tab:overall_performance}
   \vspace{-1mm}
\end{table*}

\begin{table*}[t]
\caption{
Performance comparison in Terms of AUC (\%) and pAUC (\%) with Ablation Study Results on development dataset.}
\renewcommand{\arraystretch}{1.1}
\centering 
\scalebox{0.93}{
\begin{tabular}{cl|cccccccccccc|cc}
\hline
\multicolumn{2}{c|}{\multirow{2}{*}{Methods}} & \multicolumn{2}{c}{Fan} & \multicolumn{2}{c}{Pump} & \multicolumn{2}{c}{Slider} & \multicolumn{2}{c}{Valve} & \multicolumn{2}{c}{ToyCar} & \multicolumn{2}{c|}{ToyConveyor} & \multicolumn{2}{c}{Average}  \\ \cline{3-16} 
\multicolumn{2}{c|}{}                         & AUC  & pAUC  & AUC  & pAUC  & AUC  & pAUC  & AUC  & pAUC  & AUC  & pAUC  & AUC  & pAUC  & AUC  & pAUC  \\ \hline
\multicolumn{2}{c|}{STgram(ArcFace)}      & 89.31 & 83.71 & 93.54 & 85.08 & 99.58 & 97.82 & 99.18 & 96.45 & 95.15 & 90.06 & 73.62 & 62.01 & 91.73 & 85.86  \\
\multicolumn{2}{c|}{TFSTgram(ArcFace)}    & 95.02 & 89.01 & 94.44 & 86.74 & 99.61 & 97.94 & 99.21 & 96.81 & 96.54 & 90.71 & 75.80 & 63.31 & 93.44 & 87.42 \\
\multicolumn{2}{c|}{STgram(OS-SCL)}        & 96.74 & 87.12 & 94.83 & 87.12 & 99.53 & 97.55 & 99.53 & 97.74 & \textbf{96.67} & 90.39 & 80.54 & 69.23 & 94.63 & 89.34 \\
\multicolumn{2}{c|}{TFSTgram(OS-SCL)}      & \textbf{98.41} & \textbf{94.50} & \textbf{96.25} & \textbf{88.74} & \textbf{99.63} & \textbf{98.04} & \textbf{99.85} & \textbf{99.23} & 96.55 & \textbf{91.17} & \textbf{83.56} & \textbf{69.87} & \textbf{95.71} & \textbf{90.23} \\ \hline
\end{tabular}
}
\vspace{-0.3cm}
\label{tab:ab}
\end{table*}

\begin{table}[t]
\centering
\caption{
Performance comparison in terms of mAUC (\%) on the test data of the development dataset.
}
\renewcommand{\arraystretch}{1.1} 
\scalebox{0.93}{
\begin{tabular}{c|cccc}
\hline
\textbf{Method} & \makecell{STgram-MFN \\ \cite{liu2022anomalous}} & \makecell{CLP-SCF\\ \cite{guan2023anomalous}} & \textbf{\makecell{Log-Mel \\ (OS-SCL)}} &\textbf{\makecell{TFSTgram \\ (OS-SCL)}}
\\ \hline
Fan         & 81.39 & 88.27 & 86.00 & \textbf{93.39} \\ 
Pump        & 83.48 & 87.27 & 89.43 & \textbf{94.17} \\ 
Slider      & 98.22 & 98.28 & 97.92 & \textbf{98.61} \\ 
Valve       & 98.83 & \textbf{99.58} & 99.40 & 99.57 \\ 
ToyCar      & 83.07 & 86.87 & \textbf{90.11} & 85.56 \\ 
ToyConveyor & 64.16 & 65.46 & 72.61 & \textbf{76.09} \\ \hline
Average     & 84.86 & 87.62 & 89.24 & \textbf{91.23}
 \\ \hline
\end{tabular}
}
\vspace{-0.3 cm}
\label{tab:mauc}
\end{table}

\begin{table}[thbp]
    \caption{Performance comparison of Large Pre-trained Models on development and evaluation dataset.
}
    \centering
    \fontsize{7pt}{10pt}\selectfont  
    \begin{tabular}{@{}lccccc@{}}
    
        \hline
        \textbf{Method} & \textbf{Size} & \makecell{\textbf{Development set} \\ \textbf{Mean}} & \makecell{\textbf{Evaluation set} \\ \textbf{Mean}} & \makecell{\textbf{All} \\ \textbf{Mean}} \\
        \midrule
        AnoPatch(WavLM) \cite{jiang24c_interspeech}             & 316M & 84.78         & 86.05 & 85.42 \\
        AnoPatch(ATST) \cite{jiang24c_interspeech}           & 85M & 85.12          & 88.29 & 85.71 \\
        AnoPatch(AST) \cite{jiang24c_interspeech}           & 86M &85.19          & 86.23 & 86.19 \\
        
        No. 1 \cite{Giri2020}              & 2M      & 86.27          & 89.77          & 88.02 \\
        
        AnoPatch(BEATs) \cite{jiang24c_interspeech}           & 90M & 90.93          & \textbf{94.28} & 92.58 \\

        \hline
        \textbf{Log-Mel(OS-SCL)}        & \textbf{0.88M} & 91.53          & 93.27          & 92.40 \\
        \textbf{TFSTgram(OS-SCL)}   & 1.38M  & \textbf{92.96} & 93.22          & \textbf{93.09} \\
        \hline
    \end{tabular}
    \label{lp}
\end{table}

\begin{table}[t]
\centering
\caption{
Performance comparison in AUC of different FPH reductions on the development dataset.
}
\renewcommand{\arraystretch}{1.1} 
\scalebox{0.88}{
\begin{tabular}{c|ccccccc}
\hline
\textbf{Reduction} & w/o FPH & 128 & 64 & 32 & 16 & 8 & 4 \\ \hline
Fan         & 98.22 & 97.26 & 98.41 & 98.37 & \textbf{98.52} & 85.65 & 50.19 \\ 
Pump        & 94.45 & 93.19 & \textbf{96.25} & 95.97 & 94.45 & 79.56 & 50.95 \\ 
Slider      & 99.57 & 99.61 & \textbf{99.63} & 99.61 & 99.62 & 95.34 & 39.65 \\ 
Valve       & \textbf{99.93} & 99.73 & 99.85 & 99.64 & 99.67 & 97.20 & 50.51 \\ 
ToyCar      & \textbf{96.97} & 96.27 & 96.55 & 95.98 & 96.76 & 92.62 & 51.20 \\ 
ToyConveyor & 80.54 & 76.75 & \textbf{83.56} & 83.21 & 78.29 & 66.95 & 47.33 \\ \hline
Average     & 94.94 & 93.80 & \textbf{95.71} & 95.46 & 94.55 & 86.22 & 48.30 \\ \hline
\end{tabular}
}
\vspace{-0.3 cm}
\label{fph}
\end{table}

\subsection{Experimental Setup}
\noindent\textbf{\textit{Dataset}}
For evaluation, we utilized the DCASE 2020 Challenge Task 2 development dataset \cite{Koizumi_DCASE2020_01}, which includes the MIMII \cite{Purohit_DCASE2019_01} and ToyADMOS \cite{Koizumi_WASPAA2019_01} datasets, along with an additional dataset. The MIMII dataset comprises four machine types (Fan, Pump, Slider, and Valve), while the ToyADMOS dataset includes two machine types (ToyCar and ToyConveyor). The training data from the development and additional datasets together form the final training set, which includes 41 different machine IDs. For evaluation, test data from the development and evaluation datasets were used. We did not use the latest DCASE datasets because these versions focus on domain adaptation and generalization tasks. Our focus is on addressing the frequent false alarm issue that arises when normal samples from different machine IDs are very similar.

\noindent\textbf{\textit{Implementation Details}} 
We trained a single MobileFaceNet \cite{chen2018mobilefacenets} model for all machine types, with a batch size of 64 and 300 training epochs. The initial learning rate was set to 0.0001, optimized with the AdamW optimizer \cite{Loshchilov2017DecoupledWD}, and dynamically adjusted using a cosine annealing strategy. Additionally, we employed the Exponential Moving Average (EMA) \cite{NIPS2017_68053af2} method to smooth the model parameters. During the feature extraction phase, all feature dimensions were set to 128×313, and the size of the embedding vector was 128. For the Arcface \cite{deng2019arcface} parameters, the scale factor \(s = 30\). The mixup \cite{zhang2018mixup} coefficient was sampled from a Beta distribution with parameters both set to 0.5. For models using only Log-Mel features, \(m = 0.7\) and Leaky ReLU was used as the activation function for FPH; for TFSTgram features, \(m\) was set to 0.4, with ReLU \cite{relu} as the activation function. The temperature coefficient \(\tau\) for the noise-supervised contrastive loss was set to 0.02, and the FPH compression dimension \(R\) was set to 64.

\noindent\textbf{\textit{Evaluation Metrics}}
For performance evaluation, we employed AUC, partial AUC (pAUC), and minimum AUC (mAUC) metrics. pAUC is a metric that assesses binary classification performance within a specific false positive rate range [0, p], where \(p = 0.1\) \cite{Koizumi_DCASE2020_01}. mAUC represents the worst detection performance among machines with different IDs \cite{dohi2021flow}, used to evaluate the lower bound of detection performance.

\subsection{Performance Comparison}
Table \ref{tab:overall_performance} compares our proposed method with others, demonstrating that it achieves the best performance in terms of average AUC and pAUC, with improvements of 1.63\% and 1.22\% over the state-of-the-art TASTgram (NAMix) \cite{10447764}. The performance improvement is consistent across most machine types, except for valve data. Notably, on the ToyConveyor dataset, we achieved detection performance far exceeding existing state-of-the-art methods, indicating that our framework's trained model has better discriminative ability for similar samples across different machine IDs. For CLP-SCF \cite{guan2023anomalous}, which uses two-stage contrastive learning, the fine-tuning phase may weaken the contrastive effect, which is why we adopted a one-stage approach. Moreover, using supervised contrastive learning without noise does not effectively distinguish similar samples from different machines, proving that appropriate noise and feature perturbation are the best solutions to this issue compared to the state-of-the-art CLP-SCF. Traditionally, the Log-Mel spectrogram is thought to be designed based on human auditory perception \cite{1163420}, potentially filtering out high-frequency components. In the mAUC metric (Table~\ref{tab:mauc}), using only the Log-Mel spectrogram improved by 4.38\% compared to STgram \cite{liu2022anomalous}, demonstrating that machine anomaly sound detection does not seem to rely on high-frequency components.

We also compared several large pre-trained models \cite{jiang24c_interspeech} on the DCASE2020 dataset with multiple machine IDs. Table~\ref{lp} shows the arithmetic mean of AUC and pAUC for all machines on both the development and evaluation sets. Compared to the BEATs \cite{pmlr-v202-chen23ag} model with 90M parameters, our method improved by 2.03\% on the development set. Although the performance on the evaluation set was slightly lower, there was an overall improvement of 0.51\%. Notably, we achieved this performance enhancement with only 1.38M parameters, making our method more advantageous for model deployment in real industrial environments when performance differences are minimal.

Finally, to verify the roles of TFSTgram, OS-SCL, and FPH, we conducted ablation experiments in Table~\ref{tab:ab} and Table~\ref{fph}. First, applying OS-SCL to STgram \cite{liu2022anomalous}, Table \ref{tab:ab} shows that this framework significantly improved STgram performance across multiple machines, particularly on the ToyConveyor dataset, demonstrating OS-SCL strong ability to distinguish similar samples of different IDs. The application of TFSTgram also enhanced detection performance for several machines. Next, we explored the importance of the feature perturbation mapping head. The ablation results (Table~\ref{fph}) indicate that removing the feature perturbation mapping head significantly decreased detection performance for some machines, suggesting that appropriate feature perturbation aids in anomaly detection. Additionally, we verified the impact of perturbation degree on detection results by setting different compression dimensions. When the compression dimension was continuously reduced, detection results began to deteriorate, as restoring to the original dimension becomes more challenging, leading to greater deviations and perturbations.

\section{Conclusion}
In this paper, we propose a one-stage noise-supervised contrastive learning (OS-SCL) and feature perturbation to optimize decision boundaries. The proposed TFgram, which focuses on both the specific time and frequency axes, significantly addresses the false alarm issues of similar samples from different machine IDs in machine ID-based self-supervised ASD systems. Furthermore, we achieved superior performance and excellent stability using only Log-Mel, far surpassing the performance of temporal feature fusion, demonstrating that machine anomalous sound detection may not rely on high-frequency components. Finally, we compared the latest large pre-trained models, showing that our method not only excels in detection performance but also has a minimal parameter count, making it suitable for practical industrial production environments, achieving state-of-the-art performance on the DCASE 2020 Task 2 dataset.

\clearpage
\bibliographystyle{IEEEtran}
\bibliography{IEEEexample}


\end{document}